# Optical Astrometry of Accreting Black Holes and Neutron Stars: Scientific Opportunities

John A. Tomsick (jtomsick@ssl.berkeley.edu)
UC Berkeley/Space Sciences Laboratory

Andreas Quirrenbach
University of Heidelberg

Shrinivas R. Kulkarni
Caltech

Stuart B. Shaklan & Xiaopei Pan
JPL

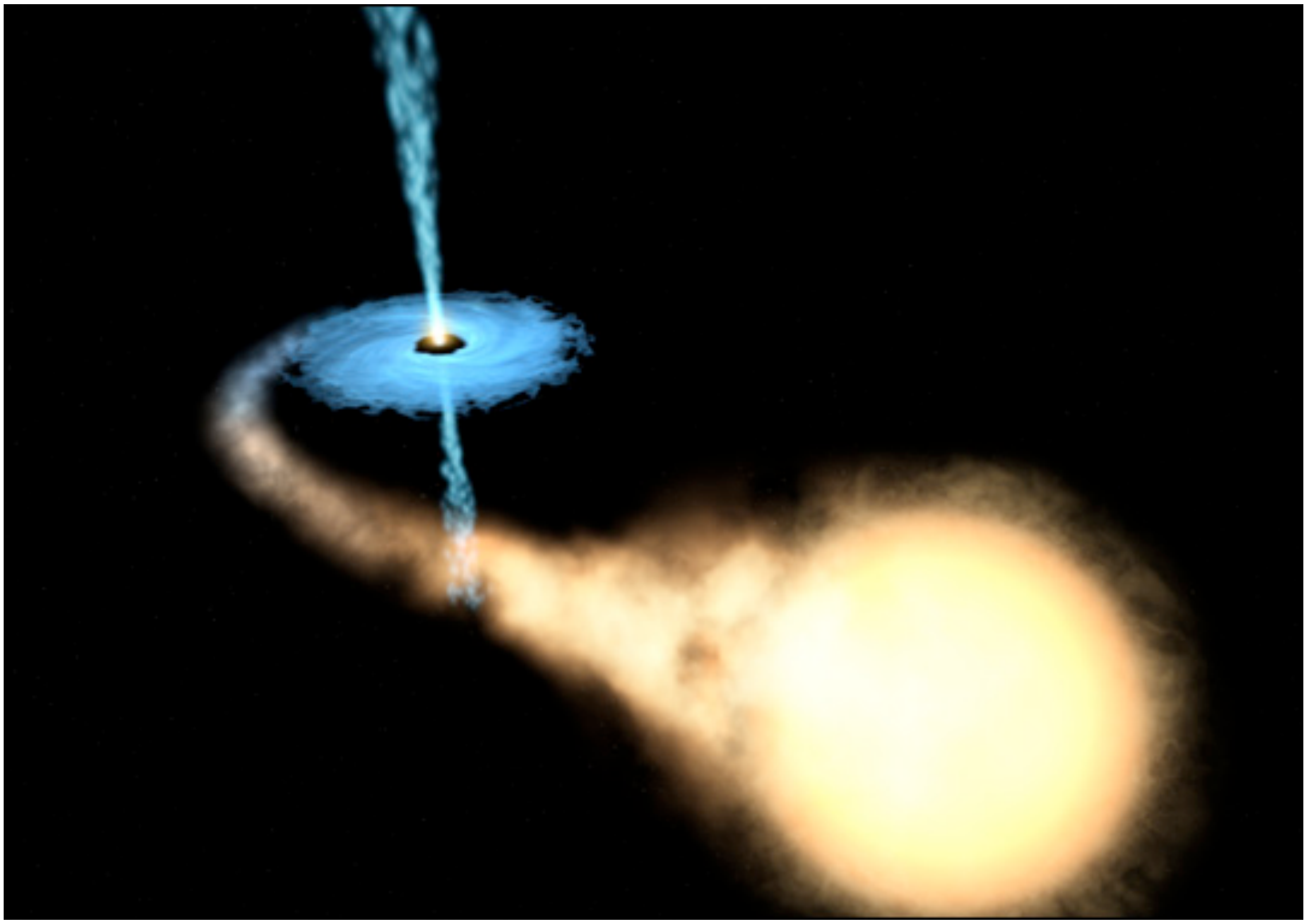

## Abstract

The extreme conditions found near black holes and neutron stars provide a unique opportunity for testing physical theories. Observations of both types of compact objects can be used to probe regions of strong gravity, allowing for tests of General Relativity. Furthermore, a determination of the properties of matter at the remarkably high densities that exist within neutron stars would have important implications for nuclear and particle physics. While many of these objects are in binary systems ("X-ray binaries"), where accreting matter from the stellar companion provides a probe of the compact object, a main difficulty in making measurements that lead to definitive tests has been uncertainty about basic information such as distances to the sources, orientation of their binary orbits, and masses of the compact objects. Optical astrometry at the microarcsecond level will allow for accurate determinations of distances and proper motions with precisions of a few percent and will provide an opportunity to directly map X-ray binary orbits, leading to measurements of orbital inclination and compact object masses accurate to better than 4%. Using astrometry at this unprecedented accuracy, which would be enabled by the Space Interferometry Mission ("SIM Lite"), will lead to breakthroughs in the investigations of the physics of black holes and neutron stars.

## Introduction

Accreting black holes and neutron stars in the Galaxy provide unique opportunities for understanding the physics of accretion, relativistic jets, and matter at nuclear densities in detail as well as providing important information about supernovae and the endpoints of stellar evolution. A primary motivation for studying X-ray binaries is that this is where the strongest gravitational fields are found (Psaltis 2008), providing an opportunity to test General Relativity (GR) in the regime of strong gravity. In addition to the great interest that these objects hold for gravitational studies, determining the properties of neutron star interiors is important for nuclear and particle physics. Although current measurement techniques still do not strongly constrain these properties, it is thought that the densities in neutron star cores may be as much as an order of magnitude above nuclear densities (Lattimer & Prakash 2007), making this the location of the densest material in the universe and making it critical to determine what form matter takes at these densities. While particle physics experiments can be used to study dense matter at high temperatures, neutron stars probe a unique region of the high-density parameter space near zero temperature (see Figure 1 and Rho 2000). Other extreme aspects of X-ray binaries include the interaction of accreting matter with very high neutron star magnetic fields as well as the relativistic jets that are often seen streaming away from neutron stars and black holes even though the details of their production are yet to be fully understood.

Precision optical astrometry can be used to obtain measurements of the quantities that are currently the most difficult to obtain. Determinations of the source luminosities, mass accretion rates, radii of neutron stars, sizes of accretion disks, and jet size-scales and velocities all depend on knowing the distances to these systems. While reliable distances to most X-ray binaries have been very difficult to obtain with the current techniques that depend on theoretical luminosities of optical companions or X-ray bursts, optical astrometry will provide parallax distance measurements. Astrometry can also be used to map binary orbits, allowing for a precise determination of the masses of black holes and neutron stars. Typically, the binary inclination is the largest contributor to the error on the compact object mass. While current measurements of the binary inclination of X-ray binaries depend on modeling the size and shape of the optical companion, optical astrometry can provide a direct measurement of this quantity. While distances and masses will be a major step forward, astrometry will also provide a measurement of the proper motions of the systems, helping to determine their birthplaces as well as answering the question of whether supernovae (or even "hypernovae") are required for the formation of black holes.

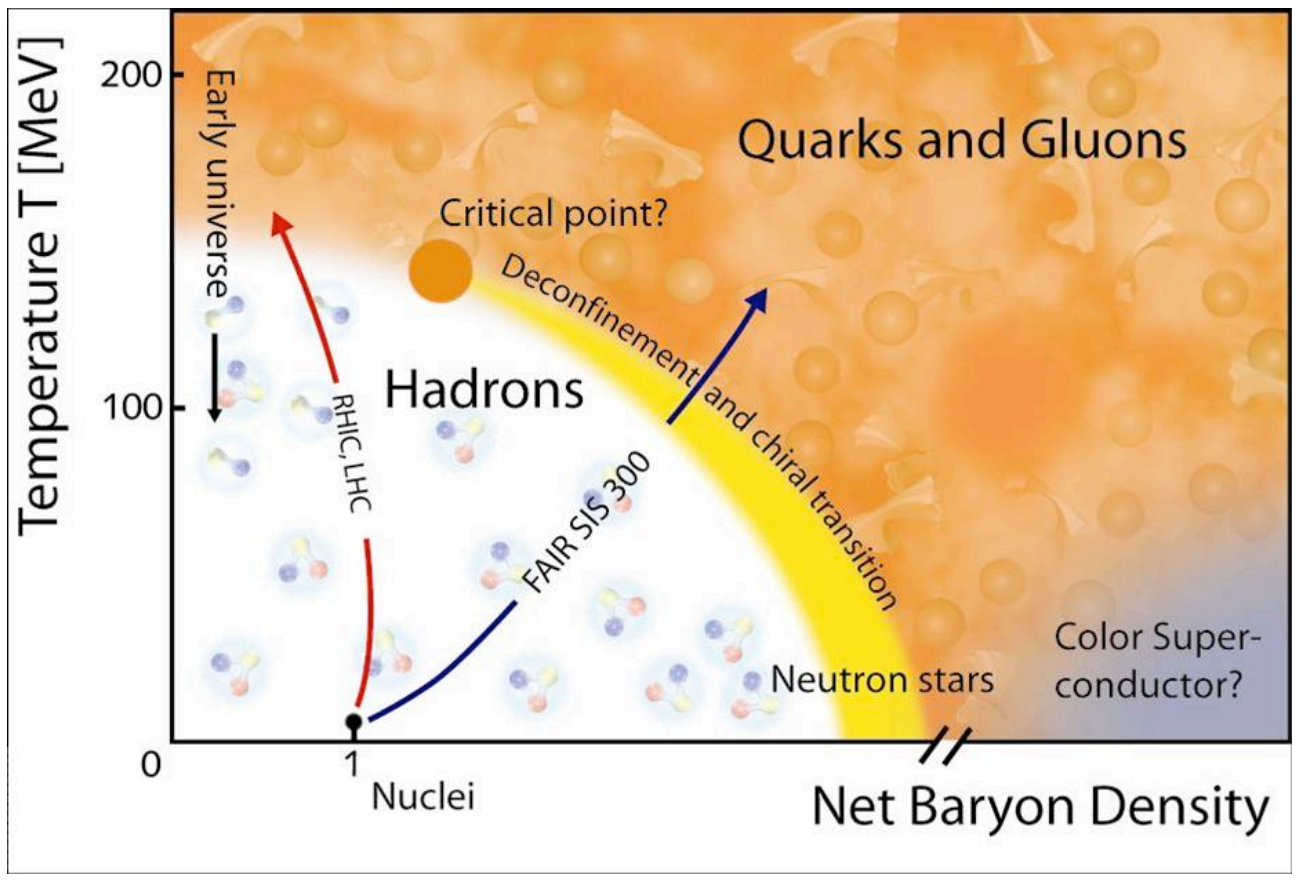

**Figure 1:** Temperature-density phase diagram for matter in different environments. Note the unique part of phase space occupied by neutron stars (adapted from Rho 2000).

The scientific topics that we emphasize here represent well-established goals in studies of compact objects. The "astrophysics challenges" that were listed in the 2000 Decadal Review (Panel Report #1; Blandford et al.) include "measure accurately the variation of neutron star radii with mass" and "form an indirect image of the flow of gas around a black hole," which mirror the first two topics discussed below ("inside neutron stars" and "probing strong gravity"). While much progress has been made on these topics in the past decade, it is clear that accurate measurements of fundamental system parameters (masses, distances, inclinations, etc.) are key for providing real tests of physical theories. We emphasize that, while the science goals we discuss are well-established, using optical astrometry to achieve these goals by making precise measurements of system parameters represents a new approach and provides new opportunities for using X-ray binaries as physics laboratories.

## Inside Neutron Stars

Although the inside of a neutron star is certainly hidden from direct viewing, significant constraints on the composition can be obtained by measuring masses and radii of neutron stars. This is because the pressure-density relationship (i.e., equation of state, EOS) that is theoretically calculated for various neutron star compositions directly predicts the star's mass-radius relation. Lattimer & Prakash (2001) discuss the different types of matter that may be inside neutron stars, including normal matter (neutrons and protons) as well as exotic matter such as hyperons, kaon condensates, and quark matter. Constraining the composition of neutron star interiors has important implications for fundamental physics such as understanding the nature of strong force interactions at high densities and determining if strange quark matter is the ultimate ground state of matter. Figure 2 illustrates the large differences in the mass-radius relationships for the different EOSs.

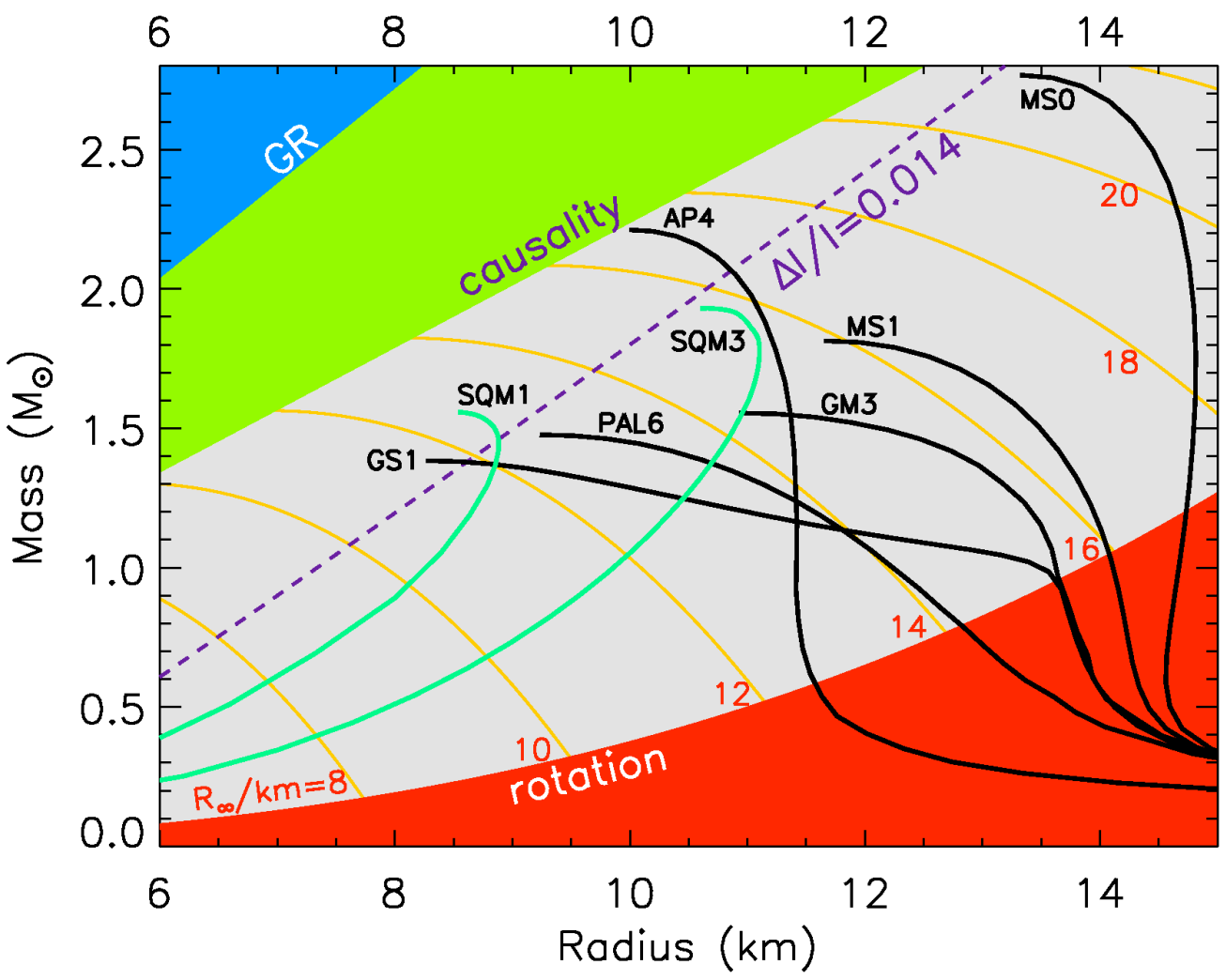

**Figure 2:** Neutron star mass-radius relationships for different equations of state (from Lattimer & Prakash 2004). The grey portion of the parameter space is allowed and includes quark matter EOSs (green lines labeled SQM1 and SQM3), EOSs with just neutrons and protons (MS0, MS1, AP4, and PAL6), and EOSs with hyperons and kaons (GM3 and GS1, respectively).

Astrometry of neutron star X-ray binaries can place significant constraints on the EOS. Each EOS predicts a maximum neutron star mass, so an accurate mass measurement for even a single neutron star above the maximum value predicted by an EOS would rule out that EOS. Although Thorsett & Chakrabarty (1999) found that many neutron stars have masses that are close to the canonical value of 1.4 solar masses, which allows essentially all EOSs, more recent observations have shown evidence for more massive neutron stars with median mass measurements in the 1.8-2.8 solar mass range (Barziv et al. 2001; Clark et al. 2002; Freire et al. 2008). A more recent summary of the mass measurements of X-ray binaries and radio pulsars is shown in Figure 3. It is clear that several of the X-ray binaries show evidence for masses in excess of 1.4 solar masses. Confirming these high neutron star masses by reducing the uncertainties would lead immediately to ruling out a large fraction of the proposed EOSs. In some cases, microarcsecond astrometry is expected to lead to neutron star mass measurements that are good to 4% or better (Tomsick et al. 2005).

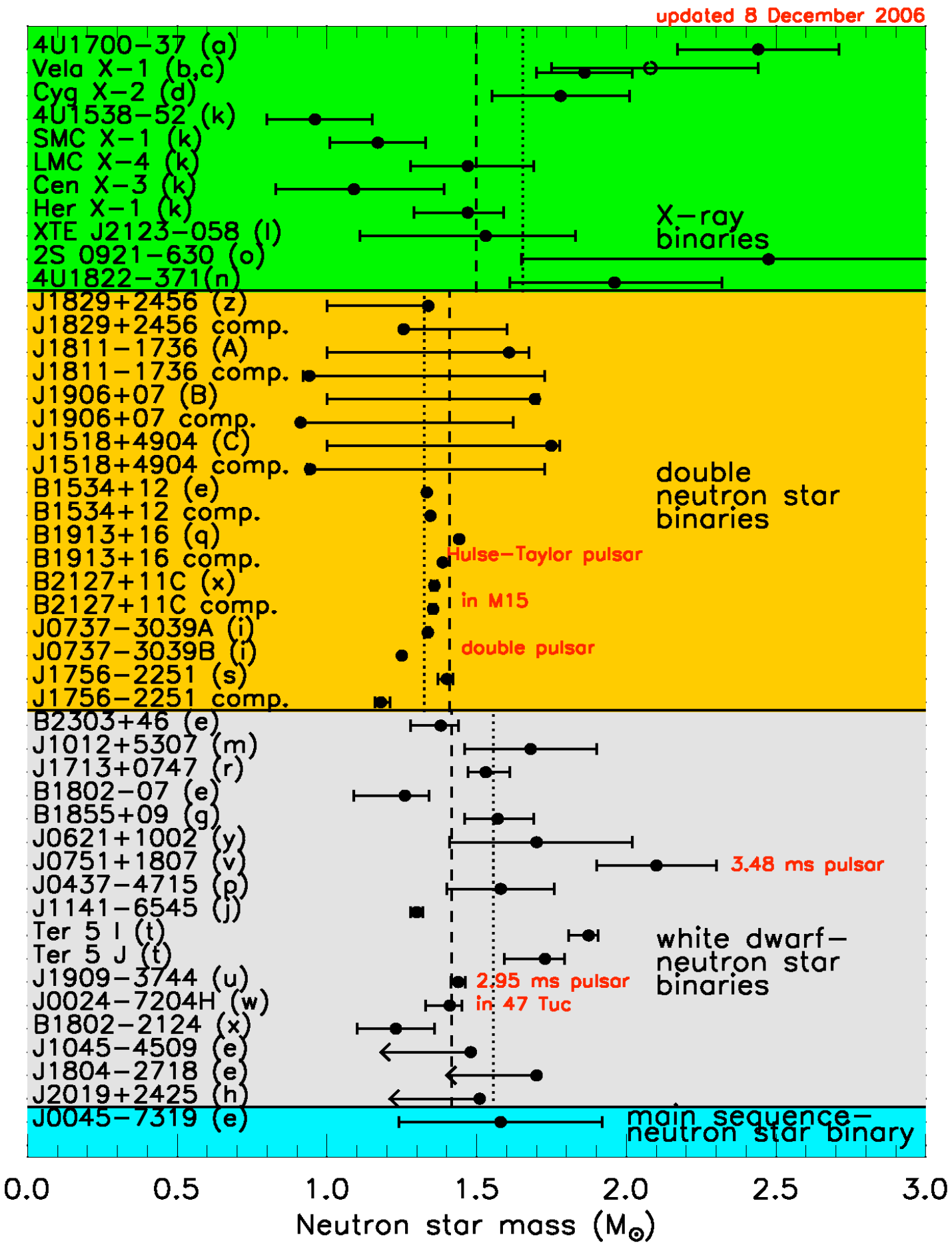

**Figure 3:** Neutron star mass measurements for X-ray binaries and radio pulsars as of 2006 December (from Lattimer & Prakash 2007). The X-ray binary measurements have not changed since the publication of this plot. However, the radio pulsar PSR J0751+1807 now has a much lower mass estimate of 1.26 solar masses (Nice et al. 2008). On the other hand, radio measurements of other pulsars indicate the possibility of high neutron star masses (Freire et al. 2008).

The parallax distances to neutron star X-ray binaries that can be obtained via astrometry will also contribute to constraining the neutron star EOS by improving measurements of the radii of neutron stars. There are at least two radius determination techniques that use X-ray measurements for which distances are a primary source of uncertainty. First, when X-ray transient systems are at their lowest flux levels, the X-ray emission is dominated by blackbody emission from the neutron star surface (Rutledge et al. 2002, Lattimer & Prakash 2007).

Thermal X-ray emission is also seen from the entire surface of the neutron star when actively accreting neutron star systems undergo X-ray bursts. Given that the radiation is thermal, one can use Planck's law to compute the area of the emitter, but to do so, one needs the distance to convert the observed brightness into intrinsic luminosity. While this is a promising technique for measuring neutron star radii, source distances are the primary uncertainty (Galloway et al. 2008). Currently, in many cases, distances to X-ray binaries are only known to a factor of two, but microarcsecond astrometry can greatly improve this situation by measuring distances to 1-6% accuracy for dozens of X-ray binaries (Tomsick et al. 2008).

## Probing Strong Gravity

Both distance and orbital measurements of X-ray binaries are important for probing the strong gravitational fields near black holes and neutron stars, potentially allowing for tests for General Relativity. One GR prediction is that black holes should, by definition, have an event horizon, i.e., a geometrical boundary from which not even light can escape. Observational evidence for this comes from a comparison of the X-ray luminosities of black hole and neutron star transients when they are at their lowest flux levels, presumably with a very low level of mass accretion. The most up-to-date results show good evidence that the black hole systems are, in fact, less luminous than neutron star systems (Narayan & McClintock 2008), and a very likely interpretation for this difference is that some of the accretion energy is advected through the black hole event horizon (Narayan et al. 1997). Still, it is notable that although this result rests on luminosity measurements, not one of the systems used to obtain the result has a direct distance measurement. Thus, obtaining parallax distances for these systems is very important.

A second GR prediction that is being tested is that black hole and neutron star accretion disks should have an innermost stable circular orbit (ISCO). The implication of an ISCO is that since there are no stable orbits within some radial distance from the compact object, the accretion disk should be truncated at that radius. One technique that is being used to measure the location of the ISCO in black hole and neutron star systems relies on measuring the shape of the iron K$\alpha$ emission line, which is produced when X-rays incident on the accretion disk cause the iron in the disk to fluoresce. At the inner edge of the accretion disk, the motion of the material produces red and blue Doppler shifts, and the gravitational field produces a redshift, causing the emission line to be extremely distorted (Tanaka et al. 1995; Miller 2007). The measurement of the inner radius comes from modeling the shape of the emission line, but the line shape also strongly depends on the inclination of the accretion disk (see Figure 4). Thus, the accuracy of any ISCO measurement is decreased if inclination must be left as a free parameter during spectral fitting. By using microarcsecond astrometry to map the orbits of accreting binaries, inclinations for some systems can be measured to a precision of 2% (Pan & Shaklan 2007), essentially removing this as a free parameter in the spectral fitting and improving constraints on the location of the ISCO.

Another ISCO measurement technique that would greatly benefit from astrometry is the effort to use the thermal continuum emission from black hole accretion disks to determine the location of the inner edge of the disk. There are times when this component dominates the X-ray emission from black hole systems, and its shape is well described by thermal spectral models. Using fully relativistic models, McClintock et al. (2006) have carried out detailed analyses for several sources to find the location of the inner radius of the disk. While the work has been successful, both distance and accretion disk inclination are uncertain parameters in the interpretation of the results (Psaltis 2008). A third ISCO measurement technique depends on black hole X-ray timing (see the separate white paper by Tomsick et al.).

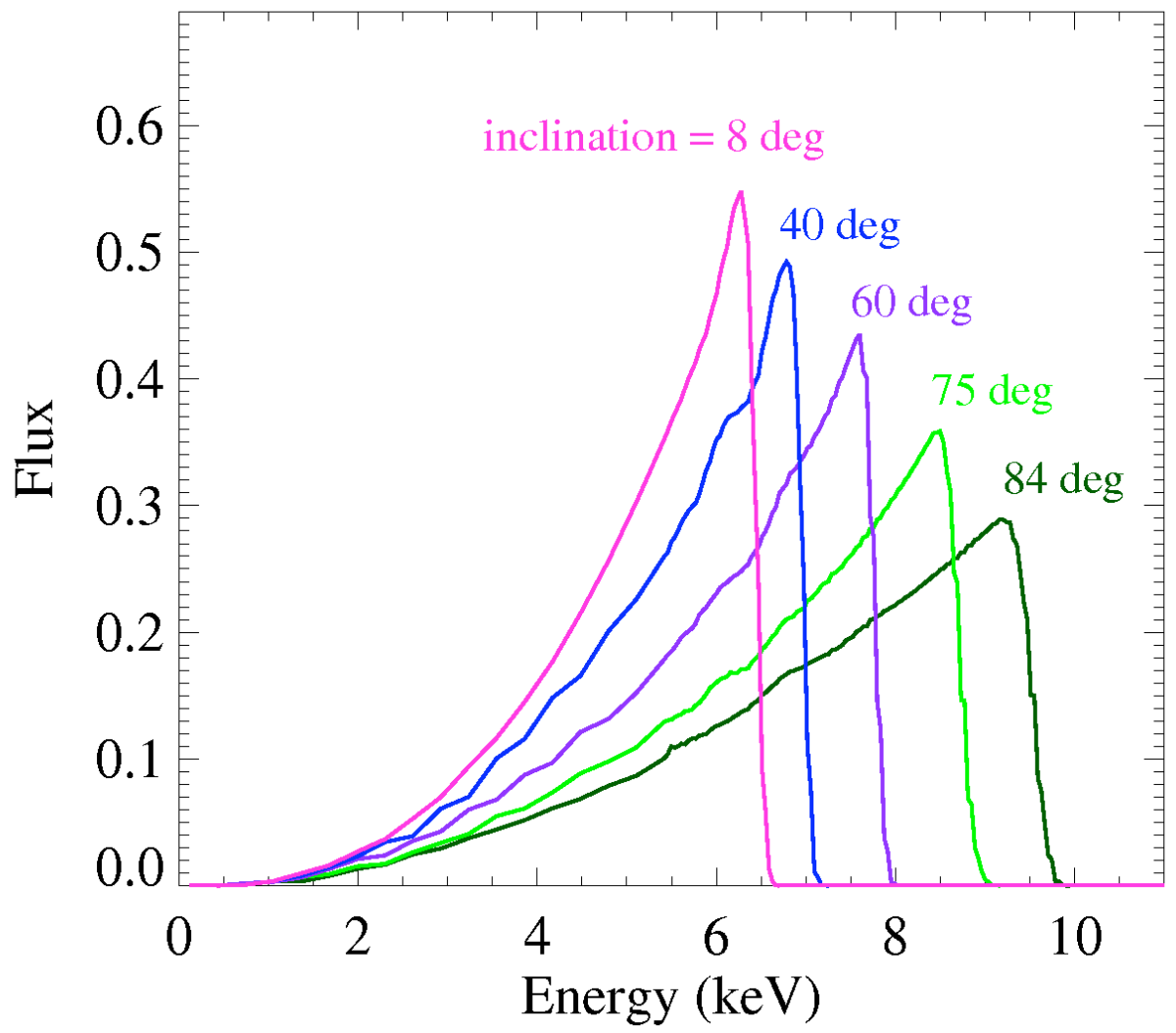

**Figure 4:** Iron line profiles demonstrating the large dependence of the line shape on binary inclination. Since the profile shapes depend on inclination as well as the location of the ISCO, a determination of the inclination via optical astrometry will improve the ISCO measurement. The profiles shown here are for an accretion disk around a maximally rotating black hole (Laor 1991).

## Birth and Evolution of Black Holes and Neutron Stars

In order to place black holes and neutron stars in the larger context of stellar life cycles, including questions related to stellar and binary evolution as well as connections to phenomena such as supernovae (or "hypernovae") and gamma-ray bursts, it is of great importance to investigate how and where these compact objects are born. By measuring the proper motions and distances to X-ray binaries, optical astrometry can help answer this question for a large number of sources. When combined with systemic radial velocities that can be obtained using spectroscopy, proper motions and distances provide a 3-dimensional space velocity that can be used to determine the object's runaway kinematics (i.e., "kick" velocity) as well as its Galactocentric orbit (Mirabel et al. 2001; Mirabel & Rodrigues 2003; Pan & Shaklan 2007). In many cases, this information can be used to test relationships between specific sources and nearby clusters of stars (e.g., Cyg X-1 and its nearby OB association) or to constrain whether a source was born in the plane of the Galaxy or out of the plane in a globular cluster.

## Capabilities Required to Achieve Science Goals

To take advantage of the science opportunities discussed above, optical astrometry will be used to measure distances and proper motions of X-ray binaries and also to map out the binary orbits. For determining the required capabilities, it is important to consider the sizes of the binary orbits as well as the optical brightnesses of potential targets. Figure 5 shows the expected orbital astrometric signatures and V-band magnitudes for a large number of neutron star and black hole X-ray binaries. While there are a few systems with signatures as large as 100 microarcseconds, measuring orbits for a significant number of X-ray binaries requires the capability to measure orbits as small as 5-10 microarcseconds for sources as faint as 15$^{th}$ magnitude. This, in turn, requires individual measurements that are accurate at the microarcsecond level. Figure 5 also shows the expected sensitivity for the Space Interferometry Mission (SIM Lite). We predict that SIM Lite will be capable of measuring orbits for about 20 neutron star and 4 or 5 black hole X-ray binaries. It is important to note that some of the most interesting targets for the science described above require microarcsecond accuracy. The current estimates of the neutron star masses for Vela X-1 and 4U 1700-377 suggest that they harbor over-massive neutron stars (also see Figure 2 above).

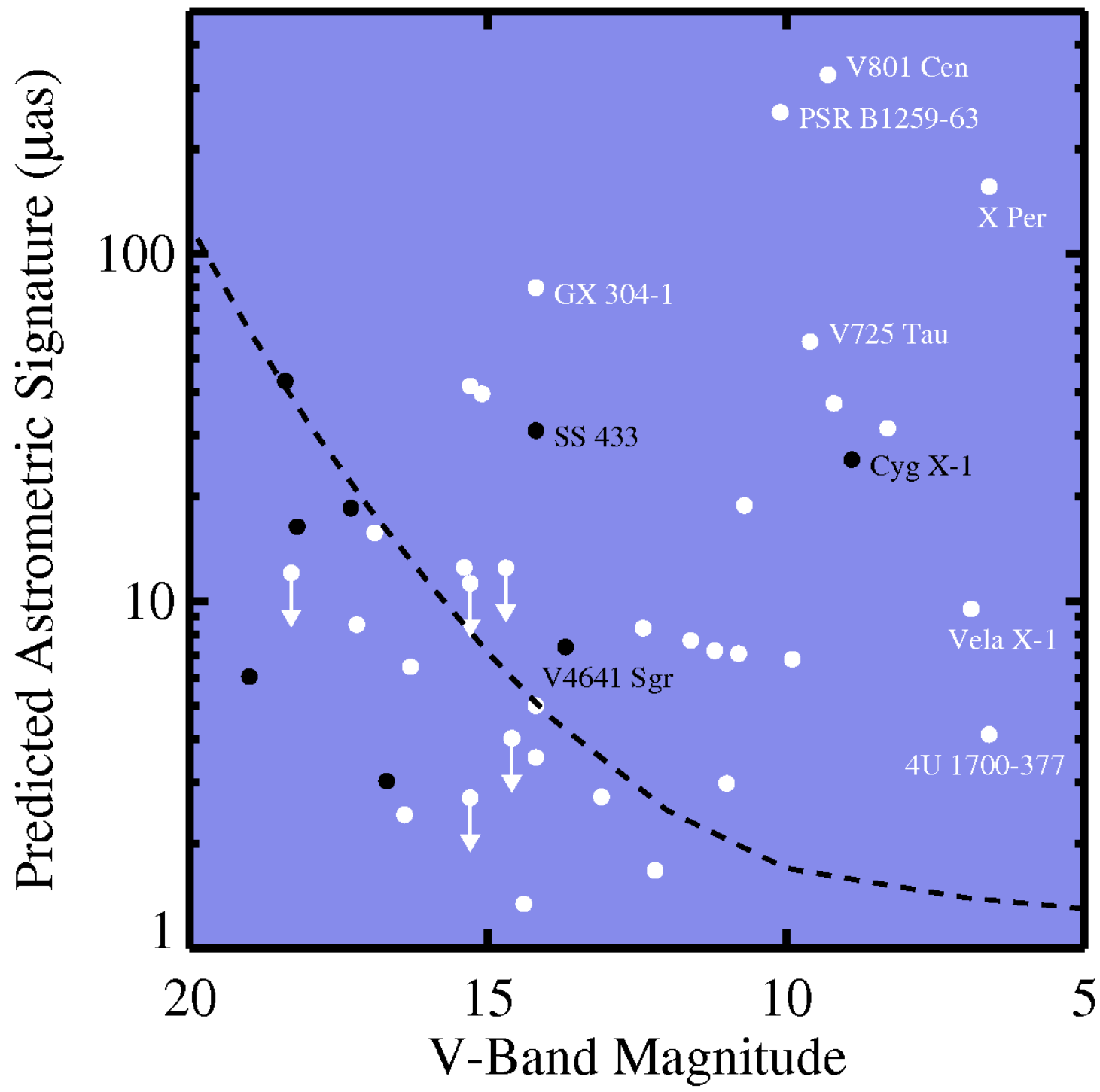

**Figure 5:** Expected astrometric signature from orbital motion vs. V-band magnitude for neutron star (white circles) and black hole (black circles) X-ray binaries. The dashed line shows the threshold for detection of orbital motion in 40 hours of SIM Lite mission time (from Tomsick et al. 2008).

For distance and proper motion measurements, 1-10 microarcsecond accuracy is sufficient for any X-ray binary in the Galaxy. However, many of the most interesting targets for such measurements, especially low-mass X-ray binaries (LMXBs), are relatively faint in the optical. As discussed in Tomsick et al. (2008) and Unwin et al. (2008), we have identified 27 LMXBs that are brighter than V=20 and would be good targets for SIM Lite. For these LMXBs, only 8 are brighter than V=17, so that a mission that can extend to 20$^{th}$ magnitude is required to obtain measurements for a significant number of targets.

Finally, we note that **Gaia**, an ESA-led optical astrometry mission (Lindegren et al. 2008), is a survey mission that will perform astrometry on large numbers of stars, but not at the precision or faintness that SIM Lite is expected to provide. Gaia can only reach (and with inferior precision) a small subset of the black holes and neutron stars that SIM Lite will reach. SIM Lite's expected microarcsecond accuracy is critical for orbital measurements, and its capability of observing faint sources is critical for distance and proper motion measurements.

## Summary and Conclusions

Optical astrometry at the microarcsecond level would provide accurate measurements of the parameters of accreting black holes and neutron stars that are currently among the most difficult to obtain. An accurate mass measurement for even one over-massive neutron star would rule out entire classes of equations of state, and determining distances to neutron star binaries is key to the measurements of neutron star radii. Both of these types of measurements will constrain the composition of neutron star cores. Astrometry will also provide great improvements in the information that we can obtain from X-ray observations of X-ray binaries. In addition to knowing their luminosities, astrometric measurements of binary inclinations would allow for more accurate measurements of the location of the ISCO. To make such measurements for large numbers of systems requires astrometric accuracy at the microarcsecond level along with the capability to observe sources as faint as 20th magnitude. The Space Interferometry Mission (SIM Lite) meets these requirements, and it would provide very exciting opportunities for studies of black holes and neutron stars.